\documentclass[conference]{IEEEtran}
\IEEEoverridecommandlockouts
\usepackage{multirow, makecell}
\usepackage[ruled,vlined]{algorithm2e}
\usepackage{cite}
\usepackage{comment}
\usepackage{amsmath,amssymb,amsfonts}
\usepackage{algorithmic}
\usepackage{graphicx}
\usepackage{textcomp}
\usepackage{xcolor}
\def\BibTeX{{\rm B\kern-.05em{\sc i\kern-.025em b}\kern-.08em
    T\kern-.1667em\lower.7ex\hbox{E}\kern-.125emX}}
\begin{document}

\title{Alice in Passphraseland: Assessing the Memorability of Familiar Vocabularies for System-Assigned Passphrases
}

\author{\IEEEauthorblockN{Noopa Jagadeesh}
\IEEEauthorblockA{\textit{Ontario Tech University}\\
Oshawa, Canada \\
jagadeeshnoopa@gmail.com}
\and
\IEEEauthorblockN{Miguel Vargas Martin}

\IEEEauthorblockA{\textit{Ontario Tech University}\\
Oshawa, Canada \\
Miguel.VargasMartin@ontariotechu.ca}
}

\maketitle

\begin{abstract}

  Text-based secrets are still the most commonly used authentication mechanism in information systems.  IT managers must strike a balance between security and memorability while developing password policies. 
  Initially  introduced  as  more  secure authentication  keys  that  people  could  recall, passphrases are passwords consisting of multiple words. However, when left to the choice of users, they tend to choose predictable natural language patterns  in  passphrases,  resulting  in  vulnerability  to  guessing  attacks. System-assigned authentication  keys  can  be  guaranteed  to  be  secure,  but  this  comes  at  a  cost  to  memorability.   In  this  study  we  investigate  the  memorability  of  system-assigned  passphrases from a familiar vocabulary to the user.  The passphrases are generated with the Generative Pre-trained Transformer 2 (GPT-2) model trained on the familiar vocabulary and are readable, pronounceable, sentence like passphrases resembling natural English sentences. Through an online user study with 500 participants on Amazon Mechanical Turk, we test our  hypothesis  -  following  a  spaced  repetition  schedule,  passphrases  as  natural  English sentences, based on familiar vocabulary are easier to recall than passphrases composed of random common words.  As a proof-of-concept, we tested the idea with Amazon Mechanical Turk participants by assigning them GPT-2 generated passphrases based on stories they were familiar with.  Contrary to expectations, following a spaced repetition schedule, passphrases as natural English sentences, based on familiar vocabulary performed similarly to system-assigned passphrases based on random common words.  
  %Our study showed that system-assigned  passphrases  based  on  random  common  words  performed  just  as  well  in terms of memorability and recall over system-assigned passphrases from a familiar vocabulary.% 

\end{abstract}


\section{Introduction}
Authentication is the process of confirming who a user claims to be. 
\begin{comment}
Authentication is a critical component of almost every computer security architecture. Everyone now has a computer and manage a  number of online accounts, such as online banking, personal e-mail, instant messaging and social networks. This means that most people need to manage specific individual credentials in order to authenticate themselves to different systems, with each system having a different rule for what is acceptable and not.
\end{comment}
Used in both corporate and personal settings, passwords have been the most popular means of authentication for many decades, despite many attempts to replace them \cite{quest}, and they continue to dominate web authentication. A typical internet user has the complex task of creating and recalling passwords for many different accounts. Most users struggle with this task, thereby adopting insecure password practices \cite{habit}\cite{factor}\cite{guess} or frequently having to reset their passwords. Allowing users to create their own passwords often leads to users picking passwords that are too easy for an attacker to guess \cite{proactive}\cite{comparison}, resulting in security breaches and loss of privacy for users. There are numerous real-world examples of password breaches \cite{data}.

To prevent users from picking passwords that are weak, of poor quality and easily guessable, system administrators adopt password-composition policies. These password composition policies force users to create more secure passwords by limiting the space of user-created passwords to exclude easily guessed passwords and thus make passwords more difficult for an adversary to guess \cite{policy}\cite{people}. Use of password-composition policies also result in a significant reduction in password reuse. However, many users struggle to create and remember their passwords under strict password-composition policies \cite{forget}. 

With a system-assigned password, a password is automatically generated by the authentication system and assigned to a user \cite{comparison}. Removing user choice and having system assigned passwords make passwords more secure. Such system-assigned passwords can be guaranteed to be sufficiently difficult to guess, however, most users find these passwords hard to remember, type \cite{study} and therefore, tend to write them down.

\begin{comment}
Multi-word passphrases may be a promising improvement over passwords. 
\end{comment}
A passphrase is a password composed of a sequence of words, intended to increase password length and therefore security, while retaining memorability \cite{pass}, however subjected to an increased rate of typographical errors. Resistance to targeted impersonation, throttled guessing, unthrottled guessing, and leaks from other verifiers \cite{quest} are some of the security advantages of system-assigned passphrases. The study \cite{correct} found that the use of system-assigned passphrases comes at a cost to memorability. Rather than memorizing, users tend to write down or otherwise store both passphrases and passwords when they are assigned by the system. It has been suggested that if people are able to pair a system-assigned passphrase with a story, it would improve the memorability \cite{comic}; however, studies indicate that even this was not a successful strategy \cite{correct}.

With the goal of improving the usability and memorability  of system-assigned passphrases, we propose in  this work a creation of  system generated passphrase based on  a familiar vocabulary to the users. As a proof-of-concept, this paper describes the results of a 500-participant study on the memorability and usability of system-assigned passphrases generated from a vocabulary familiar to the user. The passphrases are generated with the Generative Pre-trained Transformer 2 (GPT-2) model \cite{gpt2} and are readable, pronounceable, natural English sentence like passphrases. The reason for generating passphrases using GPT-2 over using phrases directly from music lyrics, movies, literature or similar public available material is to prevent dictionary attacks. The idea is to have a user-specific vocabulary, e.g., generated from user-created text content, such as emails or blog post or tweeted messages; users may also choose a pre-generated dictionary specific with their interests e.g., vocabulary  from a favorite poet or books. This user-specific vocabulary is then used to train a GPT-2 model, that generates readable, pronounceable, natural English sentence like passphrases.

Our work was motivated from a  prototype-\emph{Myphrase} \cite{myphrase} which is a multi-word password scheme that encourages users to use words that are more personal. A \emph{Myphrase} passphrase consists of  randomly chosen multiple words from a user-created/selected dictionary. Authors of \emph{Myphrase}  believed that \emph{Myphrase} passphrases may retain security advantages of random phrases without being too difficult to remember, as the dictionary is created from a familiar/personal vocabulary. However, to the best of our knowledge  this hypothesis was never tested. The main contributions of our work are as follows:
\begin{enumerate}
  \item We present a proof-of-concept approach for system-assigned passphrases that harnesses the power of  Generative Pre-trained Transformer models such as  GPT-2.
  \item We discuss the security goals to be achieved when designing a system-generated passphrase based on personal/familiar vocabulary.
   \item Through an online user study with 500 participants on Amazon Mechanical Turk, we test the hypothesis - following a spaced repetition schedule, passphrases as natural English sentences, based on familiar vocabulary are easier to recall than passphrases composed of random common words.

\end{enumerate}
\begin{comment}
For example, for a Netflix user who need to reset their password the vocabulary can come from their recently watched movie. For an Amazon account it could be based on the recently read book by the user. 
\end{comment}

%%---------------------------------------------------------------------------------------------
%%---------------------------------------------------------------------------------------------
\section{Background and Related Work}
\subsection{System-Assigned Secrets}
Shay et al. \cite{correct} experimented on using random system-assigned passphrases, by encouraging users to imagine a scene that links each word in the passphrase. In their paper they called this condition as \emph{pp-nouns-instr}. In this condition participants were assigned four nouns, randomly sampled with replacement from a dictionary containing 181 most common nouns. Participants were then asked  to construct a scene that includes all of the words in their password phrase, assuming that imagining a scene would improve the memorability of the assigned passphrase. However, their results were not very encouraging as, out of the eleven  variations on passphrase and password conditions they tested, eight conditions showed greater successful login on first try when compared to pp-nouns-instr. This makes us think how well people can construct a memorable scene when just asked to imagine one without given any directions.  

Bonneau et al. \cite{storage} demonstrated the brain’s ability to learn and later recall random full 56-bit system assigned secret.  They found that 88\% of users were able to recall their passphrase after 3 days of  completion  of  the  study, however, the training period was quite long (about 12 minutes over the course of 10 days on average). Blocki et al. \cite{repetition} provided evidence that the password strengthening mechanism of Bonneau and Schechter \cite{storage} could be improved by adopting the Person-Action-Object (PAO) story mnemonic, using a rehearsal schedule in $ 1.5 \times increasing \space intervals$. They asked their participants to imagine a story based on a photo of a scene that was shown, a user-chosen famous person from a predefined list, and a randomly selected action-object pair that served as their secret passphrase. They identified that most of the forgetting happened in the first test period (12 hours) and 89\% of users who remembered their passphrase during the first test period successfully remembered them in every subsequent round. 

Joudaki et al. \cite{reinforcing} showed how to improve the usability of system-assigned passphrases using implicit learning techniques. Their system design utilizes two implicit learning techniques: contextual cueing and semantic priming. Contextual cueing is a 2-dimensional spatial arrangement of distractors, and semantic priming is displaying semantically-related prime words. Users were assigned a 4-word system assigned passphrase and each word in the passphrase is presented in a display surrounded by 31 semantically related words. In the user study conducted, the authors were able to improve recall rates and login times for 4-word system-assigned passphrases.

Al-Ameen et al. \cite{cues} proposed providing users with a series of cues to aid recall of system-assigned secret. They called their system \emph{CuedR} and it allows users to choose from multiple cues (visual, verbal, and spatial), the one that best fit their learning process. Their pilot study showed that this method holds promise as all users recalled their phrase after 1 week within three attempts .

\subsection{GPT-2 and Language Modeling}
Language model (LM) is a machine learning model that can predict based on the words already observed in the sequence, the probability of the next word in the sequence. An example of LM are  keyboards of smartphone that suggest the next word based on what is currently typed. In February 2019, OpenAI released a paper \cite{LMGPT} describing GPT-2, a model for text-generation based on the Transformer architecture \cite{attention} and trained on huge amounts of text data. With the objective to predict the next word, given all of the previous words within some text; the GPT-2 model was trained on a dataset of 8 million web pages. This dataset is called WebText and is a  40GB dataset \cite{LM}. In the staged release OpenAI has released four flavors of GPT-2 models. The small model with 117M parameters, a 345M version, a 762M and a 1.5B model.

A neural network needs to be trained on a very large dataset and the knowledge gained from this dataset is compiled as “weights” of the network. Traditional learning is isolated and occurs purely based on the datasets and the specific tasks. No knowledge gained through traditional learning is retained which can be transferred from one model to another. Training is very expensive, both in terms of resources and time. Transfer learning is a machine learning method in which a model developed for a particular task is reused as the starting point to solve a similar problem. The Python code to download the smaller version of  GPT-2 model and the TensorFlow code to load the downloaded model and generate predictions was open-sourced on GitHub \cite{git} by OpenAI. By utilizing transfer learning and building upon OpenAI’s GPT-2 text generation model, Ressmeyer et. al. \cite{faking} trained a model for the domain-specific task of generating tweets based on a given account which proved to be more successful than using an Long Short Term Memory networks (LSTM). Budzianowski et al. \cite{hello}  adapted the pretrained GPT-2 model to multi-domain task-oriented dialogues. 

%%---------------------------------------------------------------------------------------------
%%---------------------------------------------------------------------------------------------
\begin{comment}

\section{ Leveraging Familiar Vocabulary}
The goal of our work is to determine, whether or not and to what degree, assigning a passphrase that resembles natural English sentences and are readable, pronounceable and are based on a vocabulary familiar to the user, help users better recall their assigned passphrase. For the purpose of generating a system-assigned passphrase that matches the above mentioned criterion's we first needed a vocabulary familiar to the users. For the purpose of our proof-of-concept study we obtained this familiar vocabulary from three popular books. We then trained a GPT-2 model on the familiar vocabulary, thereby generating passphrases that are readable, pronounceable, sentence like passphrases resembling natural English sentences. 
\end{comment}
%%---------------------------------------------------------------------------------------------
%%---------------------------------------------------------------------------------------------
\section{Study Design}
We conducted our two-part online user study of system-assigned authentication secret passphrases using Amazon’s Mechanical Turk (M-Turk)  framework. The study was approved by the Research Ethics Board (REB) at our institution. After participants consented to participate in the research study, we assigned each participant to a particular study condition. In the first part of the study, participants were assigned a passphrase and were asked to memorize the assigned passphrase. Six hours later, participants were invited to return back, log in using their assigned passphrase, and complete a survey. We followed a space repetition schedule of one day for six days where participants were asked to recall the assigned passphrase. In this section, we give an overview of our study design and experimental conditions.
\subsection{Study Overview}
We recruited participants through Amazon’s M-Turk. Participants needed to be at least 18 years old and we limited our data collection to participants from the United States or Canada. We compensated them 65 cents for completing the first part of the study and an additional 50 cents for completing each round of the second phase of the study (the compensation was prorated in accordance with the minimum wage in our university's location).

Our study is divided into two parts. Part 1 of the study is called as \emph{Memorization Phase} and Part 2 of the study is called the \emph{Recall Phase}. The \emph{Recall Phase} is a multiple round phase, and is composed of six rounds. We refer to the first recall round as \emph{Day$\#$1 Recall}, second recall round as \emph{Day$\#$2 Recall} until the sixth recall round \emph{Day$\#$6 Recall}. During \emph{Memorization Phase} we assigned participants a system generated passphrase. We asked them to memorize this assigned passphrase, and informed them that they will have to periodically return to the study and login to a dummy website using the assigned passphrase. The phase of the study were participants return back to check if they still remember the passphrase is the second part of the study which is referred to as \emph{Recall Phase}. During the \emph{Recall Phase} participants were asked to login to a dummy website using the assigned passphrase. During the \emph{Memorization Phase}  participants were informed that this was a memory study and they should not write down the words that we ask them to memorize. They were told that they will be paid for each completed recall phase round — even if they forgot the passphrase.

During the \emph{ Memorization phase}, we assigned participants a secret passphrase in one of two conditions, described in Subsection B. After being assigned the secret, participants were required to login to a dummy website using the assigned secret. After three unsuccessful attempts, participants were told their secret. Six hours after completing the \emph{Memorization phase} of the study, participants received an email through M-Turk asking them to return for the first round of \emph{Recall Phase} (\emph{Day$\#$1 Recall}). To begin the Recall phase, participants were asked to login using their secret to the same dummy website. After three incorrect attempts, we showed participants their secret. Once they had logged in, participants completed a survey about how they had remembered their assigned passphrase and their sentiments toward the assigned passphrase. This completes the \emph{Day$\#$1 Recall}. One day after the memorization phase, the \emph{Day$\#$2 Recall} study was conducted. It had the same procedure as \emph{Day$\#$1 Recall}, except that \emph{Day$\#$2 Recall} didn’t have a survey. Only those participants who returned to complete \emph{Day$\#$1 Recall} were sent an invitation to participate in the \emph{Day$\#$2 Recall}. This procedure continued until \emph{Day$\#$6 Recall} which is the last day for the multiple round recall phase. The participants who reach till \emph{Day$\#$6 Recall} are those who haven't dropped out from the study along any of the recall rounds.  Blocki et al. \cite{repetition} reported that much of the forgetting of system-assigned passphrases happens in the first 12 hours. This is why we chose our \emph{Day$\#$1 Recall} to be 6 hours after the \emph{Memorization Phase} after which participants have to return every day  for six consecutive days. A limitation to  \cite{repetition} is their quite long training period requiring participants to  login into a website 90 times over up to 15 days, which they did at an average rate of nine
logins per day. For this reason we wanted to keep our training period short and so we have only until \emph{Day$\#$6 Recall}.

\subsection{Conditions}
We assigned participants to one of four experimental conditions, which are summarized in Table \ref{table:1}. The conditions varied in the passphrase assigned to the participants. Participants were unable to modify their assigned passphrase or to obtain a replacement.  We required that participants enter words separated by spaces and in the same order they were assigned. All assigned passphrases were case-sensitive and they might resemble English like sentences depending on the passphrase condition.

\begin{table*}[h]
\centering
\begin{tabular}{||c c c c c||} 
 \hline
 Condition name & Length & Story & Assigned Passphrase & \\ [0.9ex] 
 \hline\hline
 Random Passphrase & 4 & NA & information lays early site &\\ 
  \hline                    

 Based on Familiar Vocabulary & 5 & Pride and Prejudice & Darcy knows my style perfectly & \\
  & 6 & The Adventures of Sherlock Holmes & Holmes investigations have always been indirect & \\
  & 7 & Alice in Wonderland & Alice was suppressed by wings of thunderstorm & \\ [1ex] 
 \hline
\end{tabular}
\caption{A summary of experimental conditions}
\label{table:1}
\end{table*}

\subsubsection{Random Passphrase Conditions}
This condition is similar to pp-sentence by Shay et al. \cite{correct} with 30 bits of entropy. Participants were assigned passphrases of the form “noun verb adjective noun,” where nouns, verbs, and adjectives are chosen from separate 181-word dictionaries. We used the same 181-word dictionary provided by Shay et al. Although these passphrases were unlikely to make semantic sense due to the random selection of words, some of them could resemble natural English sentences.

\subsubsection{Passphrase Condition based on Familiar Vocabulary}
In this condition we assign participants with a passphrase generated from a familiar vocabulary. For this we asked participants to select a story they are most familiar, from a list of three stories during the \emph{ Memorization phase}. We then assigned them with a system generated passphrase based on the vocabulary of the selected story. The three stories they chose from were \emph{ Alice in Wonderland, Sherlock Holmes and Pride and Prejudice}. These three are among the top 15 books downloaded from the Project Gutenberg library (which contains over 60,000 free eBooks) and they also have at least one movie based on the story.

We had to keep the length of passphrases based on Familiar vocabulary to be greater than four as shown in Table \ref{table:1}. This was because  passphrases less than four, from the three stories selected appeared more as random words rather than pronounceable English sentences and could not capture the style in which the story was written. We wanted to ensure that the passphrases generated from the stories not only are familiar in context of its vocabulary, but also resembles natural English sentences. 

\begin{comment}
OpenAI’s  GPT-2 is a pre-trained, transformer-based language model that can be used for various NLP tasks such as text generation, translation, and data summarization. It is trained with the goal of predicting the next word given all previous words of some text \cite{LMGPT}.
\end{comment}

By utilizing transfer learning and by fine-tuning GPT-2 text generation model, we sought to train a model for the domain-specific task of generating short stories in the style of  the selected  book. 
\begin{comment}
Fine tuning GPT-2 has now become the best practice and have demonstrated best results in several language task \cite{pretrain},\cite{patent},\cite{faking}. 
\end{comment}
In this work, our experiments are based on the 345M GPT-2 model. We forked Neil Shepperd's OpenAI’s repo \cite{neil} which contains additional code to allow fine tuning the existing OpenAI model on custom datasets. For our experiment, we built a baseline model with temperature=0.7, top\_p=0.9, top\_k=40. Top-p sampling used in combination with Top-K, can avoid very low ranked words while allowing for some dynamic selection. Higher temperatures work better (e.g. 0.7 - 1.0) to generate more interesting text. The  generated text are of length greater than 300 to accurately represent the context of the story on which the model is trained on and to generate realistic and coherent output.

Bonneau et al. \cite{storage} demonstrated user’s ability to remember randomly assigned 56-bit (encoded as either 6 words or 12 characters)  secret through spaced repetition. Bonk et al. \cite{christ} created StoryPass which asked users to create a sentence-like passphrase of at least 7 words in length. A 7-word long passphrase are difficult to be guessed due to the lack of tables for larger n-gram sizes. COCA\cite{COCA} corpus the largest, genre-balanced corpus of American English has only up to 2, 3, 4, and 5 n-grams. Figure \ref{fig:guesswork} on marginal guesswork shows that 2, 3 and 4-grams have lower marginal guesswork in bits compared to 5-grams. Keeping these factors into consideration, for our study we  choose passphrases that are 5, 6 and 7 words long.  

\begin{figure*} 
\centering
\includegraphics[scale=0.70] {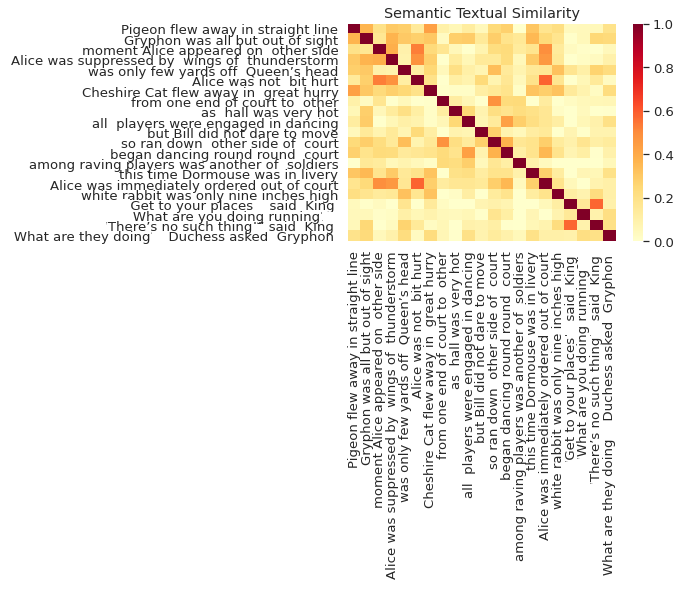}
\caption{Sentence similarity scores of passphrases in Table \ref{tab:table3} using embeddings from the universal sentence encoder \cite{USE}}
\label{fig:similarity}
\end{figure*}

To simulate the use of familiar vocabulary, we used three classic literary pieces: \emph{Alice in Wonderland} (Lewis Carroll, 1965), \emph{Pride and Prejudice} (Jane Austen, 1813), \emph{The Adventures of Sherlock Holmes} (Arthur Conan Doyle, 1892) downloaded from  Project Gutenberg library. We used GPT-2 to generate 300+ words text for each of the three books. We then cleaned the generated text to produce passphrases of 5, 6 and 7 words long. Table ~\ref{tab:table2} shows the short story generated by GPT-2 from the book \emph{Alice in Wonderland}. To produce passphrases from the GPT-2 generated short story, we performed text cleaning. To start with, each lines from the GPT-2 generated text was extracted. We then split each line into a sequence of words that forms a candidate passphrase until a punctuation - comma, colon, semicolon or period- is found. All those sequences, with word count less than 5 were discarded from the candidate pool of passphrases. The conjunction `\emph{and} ', definite article `\emph{the} ', indefinite article `\emph{a/an} ' were then removed. The word count of each candidate passphrase were then taken and those less than 5 or greater than 7 were removed from the candidate pool of passphrases. In the final step we randomly replaced personal pronouns (I, me, she, he, them, it) to character names in the book. This data cleaning operation generates passphrases of 5 , 6 or 7 words long. Table ~\ref{tab:table3} shows the passphrases generated from Table ~\ref{tab:table2}. We then ranked the generated passphrases using conditional probability of n-grams. Assuming an attacker wants to guess all n-gram combinations, the best list for an attacker with unlimited resources would be the  bigrams, because all of the word combinations that appear in bigram also appear in all the other n-grams. Before assigning a different user with a passphrase generated from the same book, we compute a passphrase similarity score to ensure the newly assigned passphrase is sufficiently different from the already assigned passphrases based on the same book. We used universal sentence encoder \cite{USE} in order to derive the semantic representation and the similarity of passphrases. We used the universal sentence encoder transformer version 4, which is available at Google Tensorflow-hub \footnote{https://tfhub.dev/google/universal-sentence-encoder/4}. For the generated passphrase in Table ~\ref{tab:table3},  Figure \ref{fig:similarity} shows the sentence similarity scores.

%%-----------------------------TABLE & FIGURES------------------------------------------------
%%-------------------------------------------------------------------------------------------------

\begin{center}
\begin{table}
    \begin{tabular}{ | p{8.0 cm} |}
    \hline
    
Alice was silent.

The Pigeon flew away in a straight line, and was just in front of it when
it came: it was all but out of sight, and the moment Alice appeared on the
other side, she was suppressed by the wings of the
thunderstorm, and was only a few yards off the Queen’s head.

Alice was not a bit hurt, and she flew away in a great hurry,
from one end of the court to the other; and as the hall was very hot, and
all the players were engaged in dancing, Alice
appeared on the fiddle: she was a good deal frightened at the
creature’s appearing, but she did not dare to move, so ran
down the other side of the court, and began dancing round and round the
court, always in a hurry: among the raving players was another
of the soldiers, and this time he was in livery, with the words “DRINK ME,”
done and ready.

“’” Said the Queen, “who demands of it so much of anything, that he’s
always ordering people to laze about in the sun!”

And Alice was immediately ordered out of the court; for, you see, as she was
only nine inches high, she could not afford to be any
more cuddly: and, by the time she had been foraging for roots and
leaves, she was getting up and walking off to the side of the court.

“Get to your places!” said the King, with a sudden emphasis. “What are
you doing running?”

“I’m a farmer,” said Alice: “I run them”

“There’s no such thing!” said the King. “You’re a serpent.”

Alice was thoroughly puzzled, and was looking down at her hands in
great curiosity. “What are they doing?” she asked the Gryphon.

“They’re doing \\ \hline
   
    \end{tabular}
    \caption{\label{tab:table2}GPT-2 generated text for \emph{Alice in Wonderland} }
    \end{table}
\end{center}

\begin{center}
\begin{table}
    \begin{tabular}{ | p{8.0 cm} |}
    \hline
    
Pigeon flew away in straight line \\
Gryphon was all but out of sight\\
moment Alice appeared on  other side\\
Alice was suppressed by  wings of  thunderstorm\\
was only few yards off  Queens head\\
Alice was not  bit hurt\\
Cheshire Cat flew away in  great hurry\\
from one end of court to  other\\
as  hall was very hot\\
all  players were engaged in dancing\\
but Bill did not dare to move\\
so ran down  other side of  court\\
began dancing round round  court\\
among raving players was another of  soldiers \\
this time Dormouse was in livery\\
Alice was immediately ordered out of court\\
white rabbit was only nine inches high\\
Get to your places said  King\\
What are you doing running\\
Theres no such thing said  King\\
What are they doing Duchess asked  Gryphon\\

\\ \hline
   
    \end{tabular}
    \caption{\label{tab:table3}Passphrases generated for \emph{Alice in Wonderland} }
    \end{table}
\end{center}
\begin{comment}

\begin{center}
\begin{table}
    \begin{tabular}{ | p{8.0 cm} |}
    \hline
    
Pigeon flew away in straight line \\
Gryphon was all but out of sight\\
moment Alice appeared on  other side\\
Alice was suppressed by  wings of  thunderstorm\\
was only few yards off  Queens head\\
Alice was not  bit hurt\\
Cheshire Cat flew away in  great hurry\\
from one end of court to  other\\
as  hall was very hot\\
but Bill did not dare to move\\
so ran down  other side of  court\\
began dancing round round  court\\
among raving players was another of  soldiers \\
this time Dormouse was in livery\\
Alice was immediately ordered out of court\\
white rabbit was only nine inches high\\
Get to your places said  King\\
Theres no such thing said  King\\
What are they doing Duchess asked  Gryphon\\

\\ \hline
   
    \end{tabular}
    \caption{\label{tab:table4}Passphrases for \emph{Alice in Wonderland} after removing non-unique tag sequence}
    \end{table}
\end{center}
\end{comment}
\section{Word selection}
Bonneau et al. \cite{ling} showed that a user tends to choose linguistically correct phrases because of which even a four words long passphrase probably has less than 30 bits of security. Their results suggest that users aren't able to choose phrases made of completely random words, but are influenced by the probability of a phrase occurring in natural language. Rao et al. \cite{grammar} investigated the effect of structural patterns on password security. They used Parts-of-Speech (POS) tagging to model the grammatical structures. Their password cracking algorithm automatically combines multiple words using their POS tagging framework to generate password guesses. They found a decrease in password search space due to the presence of grammar (also referred as tag-rule in the paper) and that this decrease can be as large as 50\% for a password of length five words. An attacker aware of such a distribution can reduce their guessing effort by enumerating values for the grammatical structure and ignoring all other structures. Building on the idea of \cite{grammar} we checked if our generated set of passphrases in Table \ref{tab:table3} follow any common grammatical structure with each other and  with the source text from which it is built.  

\begin{figure}[h]
  \centering
  \includegraphics[width=\linewidth]{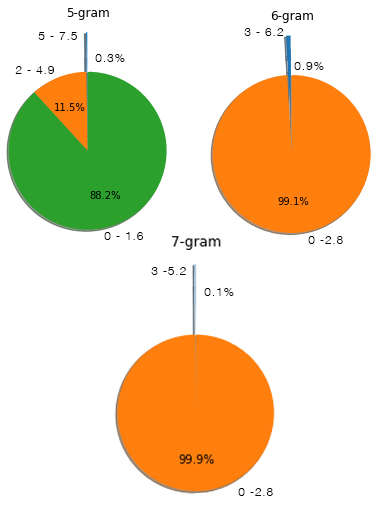}
  \caption{Tag-rules of length 5 to 7 grouped by the size of their search space (in bits) for the book \emph{Alice in Wonderland}. The search space of a tag-rule, $t_s$  is the number of word sequences it generates, which in bits is
    $log_2count(S(t_s))$. Numbers outside the pie chart indicate the range of bits. Numbers inside the
    pie indicate the percentage of tag-rules with those many bits \cite{grammar}.
    }
  
  \label{fig:tag}
\end{figure}

For the book \emph{Alice in Wonderland} after removing the punctuation symbols there are 15 POS tags. We then created 5-gram, 6-gram and 7-gram of word and POS tag pairs. Finally, we extracted the set of unique tag-rules from the word tag pair in each n-gram. In Figure \ref{fig:tag} we group the tag-rules by their search space size. We observe that majority of tag-rules have a very small or negligible search space. For example, in tag rule of length seven (7-gram), 99$\%$ of tag rules have less than 2.8 bits of strength. This is because almost 99\% of tag-rules have only one tag sequence. This implies that for 7-grams, only one sequence of words are generated by 99\% of tag rules. Because of this limitation of our chosen vocabulary, we remove from Table \ref{tab:table3} all those passphrases that don't have a tag-sequence that is either unique from each other or with the  tag-rules from n-grams. This would make sure that our generated passphrase don't follow any specific tag-rule thereby offering an entropy of $log_2 15^ 7$ = 27 bits for a seven word long passphrase. The ideal scenario is to have the tag-rules divide the password search space evenly and with every search space offering high bits of strength. The passphrase ``all players were engaged in dancing" and ``What are you doing running" from Table \ref{tab:table3} are removed as these two phrases had a tag sequence that was not unique and was similar with a tag-rule in the 6-gram and 5-gram respectively in the book \emph{Alice in Wonderland}.

The presence of semantic patterns could reduce even further the search space of passphrase. Bonneau et al. \cite{ling} showed how the presence of semantic patterns in passphrases decreases guessing efforts to an extent that even a 4-word phrase can now be less than 30 bits of security.  Bonk et al. \cite{christ} used a \emph{Passphrase Ranking Algorithm}, based on n-grams, to estimate the number of guesses it would take to crack a user's passphrase. To analyse the semantic effects of a generated passphrase  we created one dictionary for 2-gram, 3-gram, 4-gram and 5-gram corpus of \emph{Alice in Wonderland} with n-grams in each dictionary sorted by increasing order of frequency. We then computed the joint probability of each of the generated  passphrases in  using 2-gram, 3-gram, 4-gram and 5-gram along with Laplace smoothing, and generate a score for each passphrase as shown in Algorithm 1. The passphrases are then ranked in increasing order of score, with  rank one for that passphrase with minimum score. This procedure helps to identify any generated passphrases  that have a high probability of being guessed and thus can be removed from the list of generated passphrases. Table \ref{tab:table5}  shows the generated passphrases arranged in increasing order of their score. Rank 1 is for the passphrase with the minimum join probability of their word sequence. \\

\begin{algorithm} 
\SetAlgoLined
\textbf{procedure} BESTNGRAM \ PROBABILITY(C)

C = generated passphrase

a= probability of C using 2-gram

b= probability of C using 3-gram

c= probability of C using 4-gram

d= probability of C using 5-gram

score= MAX(a,b,c,d)

 \caption{Score for generated passphrase}
\end{algorithm}

\begin{center}
\begin{table}
    \begin{tabular}{ | p{8.0 cm} |}
    \hline
Rank 1: Alice was suppressed by  wings of  thunderstorm\\
Rank 2: among raving players was another of  soldiers \\
Rank 3: Cheshire Cat flew away in  great hurry\\
Rank 4: Pigeon flew away in straight line \\
Rank 5: began dancing round round  court\\
Rank 6: Alice was immediately ordered out of court\\
Rank 7: white rabbit was only nine inches high\\
Rank 8: this time Dormouse was in livery\\
Rank 9: Alice was not  bit hurt\\
Rank 10: from one end of court to  other\\
Rank 11: was only few yards off  Queens head\\
Rank 12: Gryphon was all but out of sight\\   
Rank 13: so ran down  other side of  court\\
Rank 14: Theres no such thing said  King\\
Rank 15: What are they doing Duchess asked Gryphon\\
Rank 16: moment Alice appeared on  other side\\
Rank 17: but Bill did not dare to move\\
Rank 18: Get to your places said  King\\
Rank 19: as  hall was very hot\\

\\ \hline
   
    \end{tabular}
    \caption{\label{tab:table5} Ranked Passphrases for \emph{Alice in Wonderland} }
    \end{table}
\end{center}

\section{Security Goals}
We use the book \emph{Alice in Wonderland} to illustrate security goals. The book \emph{Alice in Wonderland} has 2565 ($\approx2^{11}$) unique tokens. The expected number of guesses for a passphrase of 5 words from the book is $2^{55}$/ 2 = $2^{54}$ which is $\gg10^{6}$ and thus has negligible risk of online attack \cite{guide}. In terms of passwords/passphrases entropy can be considered as a measure of the difficulty of guessing a password/passphrase \cite{nist}. For a 5-words passphrase from  \emph{Alice in Wonderland}, the entropy in bits is $log_2 2565^ {5}$ = 56 bits.

Marginal guesswork, is defined in \cite{guesswork} as ``the optimal number of trials necessary to be guaranteed a certain chance of guessing a random value in a brute-force search". Bonneau et al. \cite{ling} used  marginal guesswork model to measure the guessing difficulty of a distribution. The marginal guesswork of X is defined as \cite{guesswork}:
\begin{equation}
w_\alpha(X) = min \Bigg\{  i \ \Bigg | \sum_{j=1}^{i} p_{[j]} \ge \alpha \Bigg\}
\end{equation}

\begin{figure}[h]
  \centering
  \includegraphics[width=\linewidth]{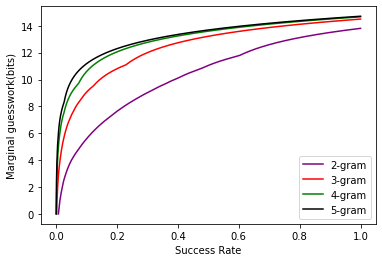}
  \caption{Marginal Guesswork for the book \emph{Alice in Wonderland} \cite{ling}.
    }
  
  \label{fig:guesswork}
\end{figure}

Figure \ref{fig:guesswork} shows the marginal guesswork for the book \emph{Alice in Wonderland}. This can determine at what point in guessing effort a certain percentage of n-grams would be guessed. The Success Rate ($\alpha$) in Figure \ref{fig:guesswork} is the probability of n-gram guessed for \emph{Alice in Wonderland}. The marginal guesswork for \emph{Alice in Wonderland} wouldn't withstand an unthrotted attack. However, we are not constructing our passphrases from n-grams of the selected book instead we are generating a new short story based on the selected book and then generating a passphrase based on that new short story. We are also ranking our passphrase based on the sequence probability as shown in Table \ref{tab:table5}. This helps us to blacklist and thereby eliminate those passphrases which are guessable according to n-grams. Our work is a proof-of-concept to a broader idea about using familiar vocabulary to improve memorability.

%%------------------------------RESULT---------------------------------------------------
%%---------------------------------------------------------------------------------------
\section{Results}
In this section, we present the results of our study.
\subsection{Study Data}
One of the primary challenges in analyzing the results from any multi-phase user study is that, some participants have to be dropped from the study because they were unable to return for one of their rounds of recall phase in a timely manner, or have left the study. It is hard to determine how many of those dropped participants would have been able to remember their passphrase, if returned. We use the following notations from \cite{repetition}. \\
\textbf{Notation}: NumSuccessfulReturned \emph{(C, i)} denote the total number of participants from study condition \emph{C} who survived (passed) through recall \emph{i} – 1 and returned for recall \emph{i}. NumRemembered \emph{(C, i)} denote the number of participants who remembered their passphrase during recall \emph{i} with < 3 incorrect attempts. NumSurvived \emph{(C, i)} denote the number of participants who also remembered their passphrase during every prior recall rounds.  NumSuccessful \emph{(i)}  denote the total number of participants  who failed recall \emph{i} – 1 but passed recall $i$.
Because the study condition \emph {C} is often clear from the context, we will omit it from the 
notations. The conditional probability that a participant remembers their passphrase during  recall \emph{i} given that they have survived through recall \emph{i}-1 and returned for recall \emph{i } is
\[ \frac{NumSurvived \emph{( i)}}{NumSuccessfulReturned \emph{(i)}} \] and is shown in Table \ref{tab:table6}. Table \ref{tab:table6} also shows how many participants who had never failed before returned in each recall rounds. \emph{i}= 0 in Table \ref{tab:table6} indicates the \emph{Memorization Phase}. We had 250 participants for Random Passphrase condition and 250 participants in the familiar vocabulary condition. \emph{i}= 1 indicates \emph{Day$\#$1 Recall} and \emph{i}= 6 indicates \emph{Day$\#$6 Recall}. Table \ref{tab:table7} summarizes the success and failure rate at each of the recall rounds. Droupout \emph{(C,i)} indicates those participants who participated in recall \emph{i}-1 but did not return for recall \emph{i}. For \emph{i} = 1 the previous phase is \emph{Memorization Phase}, denoted as  \emph{i} = 0 in Table \ref{tab:table7}.

\begin{table*}
\centering
\begin{tabular}{|l||c|c|c|c|c|c|c|} \hline\hline

      \textbf{ Recall Condition} & \textbf{i=0} & \textbf{i=1} & \textbf{i=2} & \textbf{i=3}  & \textbf{i=4}  & \textbf{i=5} & \textbf{i=6}\\
      
      \hline
      \multirow{2}{*}{Random} & 250 & 148, & 89, & 96, & 101, & 107, & 92, \\ 
      &  & 94,  & 87, & 82, & 77, &74, & 67,\\
      &  & 63.51$\%$ & 97.75$\%$ & 85.42$\%$ & 76.24$\%$  & 69.16 $\%$ & 72.83$\%$\\
      \hline
      \multirow{2}{*}{Familiar Vocabulary} & 250 & 162, & 94, & 96, & 99, & 98,& 88,\\ 
      & & 105, & 87, & 70, & 65, & 61, & 52,\\
      & & 64.81$\%$  & 92.55$\%$  & 72.92$\%$ & 65.66$\%$ &62.24$\%$ & 59.09$\%$\\
      \hline

\end{tabular}
\caption{\label{tab:table6} NumSuccessfulReturned \emph{(i)} , NumSurvived \emph{( i)} , NumSurvived \emph{( i)}$/$NumSuccessfulReturned \emph{(i)} }
\end{table*}

%%-----------------------------------------------------------------------------------------------
\begin{table*}
\centering
\begin{tabular}{ |p{1cm}||p{1.5cm}|p{2.2cm}|p{1cm}|p{1cm}|p{1cm}|p{1cm}|p{2.0cm}|}
 \hline
 \multicolumn{8}{|c|}{\textbf{Random}} \\ 
 \hline
 i = & No.of Participants & NumRemembered \emph{(i)} & Failed Login & Success Rate($\%$) & Failure Rate($\%$) & Drop out \emph{(i)} & NumSuccessful \emph{(i)} \\ 
 \hline
 0 & 250 & & & & & &\\
 1   & 148   & 94 &  54 & 63.51 & 36.49 & 102 &\\
 2 &   136  & 101   & 35 & 74.26 & 25.74 & 12 & 14\\
 3 & 130 & 106 &  24 & 81.54 & 18.46 & 6 & 11\\
4   & 122 & 110 & 12 & 90.16 & 9.84 & 8 & 10\\
 5 & 119  & 102 & 17 & 85.71 & 14.29 & 3 & 2\\
6 & 106  & 98   & 8 & 92.45 & 7.55 & 13 & 7\\
 \hline
\end{tabular}

\end{table*}

\begin{table*}
\centering
\begin{tabular}{ |p{1cm}||p{1.5cm}|p{2.2cm}|p{1cm}|p{1cm}|p{1cm}|p{1cm}|p{2.0cm}|}
 \hline
 \multicolumn{8}{|c|}{\textbf{Familiar Vocabulary}} \\ 
 \hline
 i = & No.of Participants & NumRemembered \emph{(i)} & Failed Login & Success Rate($\%$) & Failure Rate($\%$) & Drop out \emph{(i)} & NumSuccessful \emph{(i)}\\ 
 \hline
  0 & 250 & & & & & &\\
 1   & 162   & 105 &  57 & 64.81 & 35.19 & 88 & \\
 2 &   146  & 113   & 33 & 77.40 & 22.60 & 16 & 26\\
 3 & 124 & 104 &  20 & 83.87 & 16.13 & 22 & 12\\
4   & 119 & 104 & 15 & 87.39 & 12.61 & 5 & 8\\
 5 & 110  & 103 & 7 & 93.64 & 6.36 & 9 & 5\\
6 & 94  & 87   & 7 & 92.55 & 7.45 & 16 & 2\\
 \hline
\end{tabular}
\caption{\label{tab:table7} Success and Failure rate of (Top) Random Vocabulary (bottom) Familiar Vocabulary }
\end{table*}
%%------------------------------------------------------------------------------------------------
\subsection{User Sentiment}
\begin{figure*} 
\centering
\includegraphics[width=\linewidth]{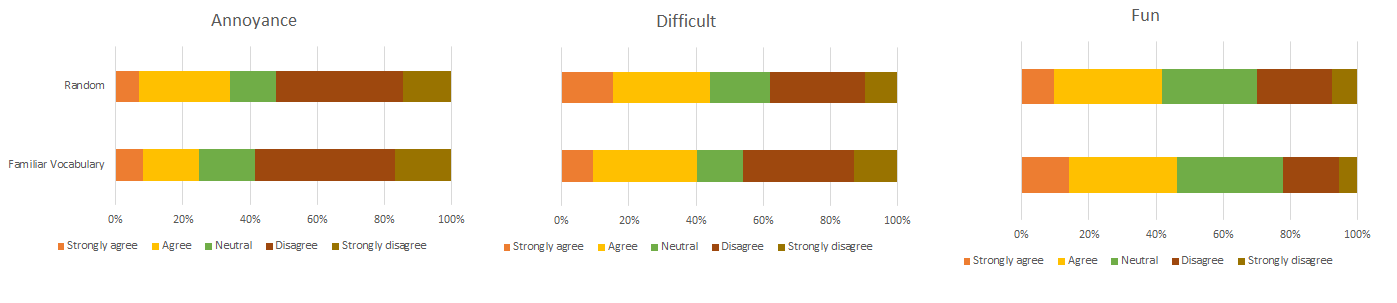}
\caption{Response data on annoyance, difficulty, and fun}
\label{fig:sentiment}
\end{figure*}

In the first round of recall phase of the study, we asked participants to indicate their agreement, from “strongly disagree” to “strongly agree,” with the statements “learning my passphrase was [annoying / difficult / fun].”
An overview of this result is shown in Figure ~\ref{fig:sentiment}.

\subsection{Story Selection}
As discussed in Section III, we  asked participants in the familiar vocabulary condition of the study to select from three stories, one story they are most familiar with. It is based on the vocabulary of this selected story we assign a passphrase to the participants. 
\begin{comment}
Participants who selected the story  \emph{Alice in Wonderland} were assigned the passphrase \emph{Alice was suppressed by wings of thunderstorm}, who selected the story \emph{Sherlock Holmes}  were assigned the passphrase \emph{Holmes investigations have always been indirect} and, for those who selected the story \emph{Pride and Prejudice}, they were assigned the passphrase \emph{Darcy knows my style perfectly}.
\end{comment}
67.90$\%$  participants selected the story  \emph{Alice in Wonderland}, 21.60$\%$  selected \emph{Sherlock Holmes} and 10.49$\%$ selected the story \emph{Pride and Prejudice}. During the first recall round we asked three questions related to the selected story, to the participants in the familiar vocabulary condition. The first question  asked was if the participants have read the chosen story or have watched a related movie. The second question was if the participants imagined a scene related to the passphrase assigned or words in the passphrase and the third question was if the imagined scene was related to the story they had chosen. 22 participants reported that they haven't read any of these three book nor watched any movie related to the these stories. Since we couldn't provide a familiar vocabulary for these 22 participants, we removed their data from our study, leaving us 162 participants for \emph{Day$\#$1 Recall}. 53.09$\%$ of participants imagined a scene  related to the passphrase assigned or words in the passphrase, out of these participants  67.44$\%$ imagined a scene related to the story they selected. 

\subsection{Typographic Error Analysis}
\begin{comment}

\begin{figure} 
\centering
\includegraphics[scale=0.55]{typo1.png}
\caption{Similarity Score between the assigned passphrase and the incorrectly typed passphrase}
\label{fig:typo}
\end{figure}
\end{comment}
Developed by Google, Bidirectional Encoder Representation from Transformers (BERT) \cite{bert} is a state of the art technique for natural language processing.  Bao et al. \cite{will_go} worked on different ways to measure similarity between embedding vectors generated by BERT, mixing Euclidean distance with cosine similarity. We use a pre-trained BERT model \cite{embedding}  to compute a dense vector representation along with cosine similarity to measure the similarity between the assigned passphrase from familiar vocabulary conditions and the failed logins.

\begin{table*}[h]
\resizebox{\linewidth}{!}{%
\begin{tabular}{|l||c|c|c|c|c|c|c|c|c|c|c|} \hline

\hline
 \multicolumn{12}{|c|}{\textbf{\ \ \ \ \ \ \ \ \ \ \ \ \ \ \ \ \ \ \  \ \ \ \ \ \ \ Number of Incorrect Passphrases and their Similarity score with Assigned Passphrase}} \\ 

 \hline
 
      \textbf{ Familiar Vocabulary} & \textbf{$>$ 95$\%$} & \textbf{95$\%$ - 90$\%$} & \textbf{89$\%$ - 85$\%$} & \textbf{84$\%$ - 80$\%$}  & \textbf{79$\%$ - 75$\%$}  & \textbf{74$\%$ - 70$\%$} & \textbf{69$\%$ - 60$\%$} & \textbf{59$\%$ - 50$\%$} & \textbf{49$\%$ - 40$\%$} & \textbf{39$\%$ - 20$\%$} & \textbf{completely dissimilar}\\
      
      \hline
      
      \multirow{1}{*}{Pride and Prejudice} & 2 & 5 & 2 & 2 & 4 &  & & & & & 3\\ 
     
      \hline
      \multirow{1}{*}{Sherlock Holmes} & 7 & 13 & 2 & 2 & 4 & 13 & 12 & 23 & 7 & & 31\\ 
    
      \hline
       \multirow{1}{*}{Alice in Wonderland} & 86 & 42 & 35 & 29 & 24 & 8 & 13 & 8 & 10 & 22 & 8 \\ 
 
      \hline

\end{tabular}}
\caption{\label{tab:table8} Similarity Score between the assigned passphrase and the incorrectly typed passphrase}
\end{table*}

For the 5-word passphrase based on the book \emph{Pride and Prejudice} there were six failed login attempts in total during the course of six recall rounds. Out of these six failed login attempts, it was just one login in which the particular participant completely forgot the passphrase and was not able to recall any one word in the 5-word passphrase with all three attempts. Since we had six failed login attempts and each login allows up to three attempts, we collected $6\times 3= 18$ incorrect ways in which the assigned passphrase was typed. Two of the incorrect passphrases had a similarity score greater than 95$\%$ with the assigned passphrase. These two passphrases are "darcy knows my style perfectly" and "Darcy matches my style perfectly" with a similarity score of 99$\%$  and 96$\%$ respectively with the assigned passphrase. We also included in Table \ref{tab:table8}, under the column `completely dissimilar' the total number of attempts in which a incorrect passphrase with no single word in the assigned passphrase was typed. 

For the 6-word passphrase based on the book \emph{Sherlock Holmes} there were 38 failed login attempts in total during the course of six recall rounds. Out of these 38 failed login attempts, there were 10 login in which the particular participant completely forgot the passphrase and was not able to recall any one word in the 6-word passphrase. In 22 of the 38 failed login attempts, participants remembered the phrase - ``Holmes investigations" from the passphrase ``Holmes investigations have always been indirect". We collected 114 incorrect passphrases for the 6-word passphrase and its similarity score with the assigned passphrase is shown in Table \ref{tab:table8}.

For the 7-word passphrase based on the book \emph{Alice in Wonderland} there were 95 failed login attempts in total during the course of six recall rounds. Out of these 95 failed login attempts, it was just one login in which the particular participant completely forgot the passphrase and was not able to recall any one word in the 7-word passphrase with all three attempts. We collected 285 incorrect unique passphrases for the 7-word passphrase and computed their similarity score with the assigned passphrase is shown in Table \ref{tab:table8}.

%%------------------------------DISCUSSION ---------------------------------------------------
%%---------------------------------------------------------------------------------------
\section{Discussion and Conclusion}
This work presents a new authentication method based on familiar vocabulary. With a user study we investigated the memorability of system-assigned passphrases from a familiar user vocabulary. The highlights of our work are presented below.  

We were able to identify a familiar user vocabulary. For conducting the experiment we chose three stories, of which we had to assume that participants would be familiar with at least one of them and would have either read a book related to one of these stories or have watched a movie related to one of the stories. Out of the participants who returned for \emph{Day$\#$1 Recall}, only 22 participants reported, to not have read a book or watched a movie related to the story they had chosen to construct the passphrase. We removed from our study the data from these 22 participants. This indicates for the remaining participants we were able to assign a system-assigned passphrase based on their familiar vocabulary. However, we acknowledge the fact that these stories might not be the most familiar and the most recent vocabulary to the participants. 

The central idea of our study is to investigate the impact of familiar vocabulary to generate passphrases that are memorable. For our study we selected three stories as this was the best way we could have known users familiar vocabulary in advance. In real-life scenarios this familiar vocabulary would be more personalized and could come from the ten books read, or from user-authored content such as sent emails and blogs.

\begin{comment}

Contrary to expectations, passphrases based on familiar vocabulary was reported to be annoying compared to random passphrases. We believe that this has to do with the length of the passphrases assigned in the two conditions. Random passphrases were  four word long whereas passphrases based on familiar vocabulary were five, six and seven words long depending on the story selected as shown in Table \ref{table:1} (please refer to Section 4.2.2 for a discussion on passphrase length). But when conducted a paired t-test on the annoyance of random and familiar vocabulary condition we find that 
%with the critical value of $\alpha= .05$ we received a probability $\emph{p}=0.5$ which indicates that % 
the annoyance of passphrases based on familiar vocabulary is not significantly high. 
\end{comment}
Passphrases based on familiar vocabulary was reported to be fun and less difficult and annoying compared to random passphrases. However, when conducted a paired t-test on the difficulty, fun and annoyance of random and familiar vocabulary condition  we find
%received a probability $\emph{p}=0.4$ and $\emph{p}=0.3$ respectively indicating %
that the difference is not significant. We believe that this has to do with the length of the passphrases assigned in the two conditions. Random passphrases were  four word long whereas passphrases based on familiar vocabulary were five, six and seven words long depending on the story selected as shown in Table \ref{table:1} (please refer to Section III for a discussion on passphrase length).

Table \ref{tab:table7} indicates a similar success rate in memorizing a familiar vocabulary passphrase and a random passphrase, even though the word lengths do not match. When conducted a paired t-test on the Success Rate of random and familiar vocabulary condition in Table~\ref{tab:table7} we find no significant difference.
%with the critical value of $\alpha= .05$ we received a probability $\emph{p}=0.2$. %
A paired t-test on the login success rate of random passphrase and
familiar vocabulary passphrase indicates that the success rate of familiar vocabulary is
not significantly greater than the random condition. The system-assigned passphrases created from a set of random words performed just as well in terms of memorability over system-assigned passphrases from a
familiar vocabulary.

In order to identify whether failed participants in the familiar vocabulary passphrase are able to memorize their passphrase over time, we identified NumSuccessful \emph{(i)} to denote the total number of participants from study condition \emph{Familiar Vocabulary} who failed recall \emph{i} – 1 but passed recall $i$. As summarized in Table \ref{tab:table9}, the participants who failed the \emph{i} – 1 recall round, are eventually memorizing their passphrase at the next recall rounds for all three stories. We also investigated from Table \ref{tab:table9} whether it is the participants who are failing in recalling the passphrase, the ones dropping out from the study. However no such relationship exists.

Table \ref{tab:table8} summarizes how similar are the incorrect passphrases typed by the participants with the assigned passphrase in the familiar vocabulary condition. There were incorrect passphrases in all three familiar condition which had no single word similar to the assigned passphrases. However, keeping that aside, for the story \emph{Pride and Prejudice} even the least similar incorrect passphrase has a similarity score of 0.77 with the actual assigned passphrase of five words. There is a clear indication that as the length of the assigned passphrase increases the participants who failed one or more recall rounds are having less similar passphrase, than to the one assigned. For the story \emph{Sherlock Holmes} there was no such incorrect passphrase that had a similarity score less than .40 with the assigned passphrase. The least similar incorrect passphrase has a similarity score of 0.41 with the actual assigned passphrase of six words. However, for the seven word passphrase based on the story \emph{Alice in Wonderland}, there were 22 incorrect passphrases that had a similarity score less than .40 with the assigned passphrase with the least similarity score of .24 for the incorrect passphrase ``alice has her tea".

\begin{table*}[h]
\centering
\begin{tabular}{ |p{1cm}||p{1.5cm}|p{1.2cm}|p{1.2cm}|p{1cm}|p{1.8cm}|}
 \hline
 \multicolumn{6}{|c|}{\textbf{Pride and Prejudice}} \\ 
 \hline
 i = & No.of Participants & Success Rate($\%$)  & Failure Rate($\%$) & Drop out \emph{(i)} & NumSuccessful \emph{(i)} \\ 
 \hline
 1   & 17  & 82.35 &  17.65 &  & \\
 2 &  16  & 87.5   & 12.5 & 1 & 2 \\
 3 & 14 & 92.86 &  7.14 & 2 & 1 \\
4   & 14 & 100 & 0 & 0 & 1 \\
 5 & 11  & 100 & 0 & 3 & NA\\
6 & 8 & 100   & 0 & 3 & NA \\
 \hline
\end{tabular}
 
\end{table*}

\begin{table*}[h]
\centering
\begin{tabular}{ |p{1cm}||p{1.5cm}|p{1.2cm}|p{1.2cm}|p{1cm}|p{1.8cm}|}
 \hline
 \multicolumn{6}{|c|}{\textbf{The Adventures of Sherlock Holmes}} \\ 
 \hline
 i = & No.of Participants & Success Rate($\%$)  & Failure Rate($\%$) & Drop out \emph{(i)} & NumSuccessful \emph{(i)}  \\ 
 \hline
 1   & 35  & 51.43 &  48.57 &  & \\
 2 &  32  & 71.88   & 28.12 & 3 & 6 \\
 3 & 28 & 78.57 &  21.43 & 4 & 2 \\
4   & 26 & 80.77 & 19.23 & 2 & 2 \\
 5 & 23  & 95.65 & 4.35 & 3 & 1\\
6 & 18 & 100   & 0 & 5 & 1 \\
 \hline
\end{tabular}
 
\end{table*}

\begin{table*}[h]
\centering
\begin{tabular}{ |p{1cm}||p{1.5cm}|p{1.2cm}|p{1.2cm}|p{1cm}|p{1.8cm}|}
 \hline
 \multicolumn{6}{|c|}{\textbf{Alice in Wonderland}} \\ 
 \hline
 i = & No.of Participants & Success Rate($\%$)  & Failure Rate($\%$) & Drop out \emph{(i)} & NumSuccessful \emph{(i)} \\ 
 \hline
 1   & 110  & 66.36 &  33.64 &  & \\
 2 &  98  & 77.55   & 22.45 & 12 & 18 \\
 3 & 82 & 84.15 &  15.85 & 16 & 9 \\
4   & 79 & 87.34 & 12.66 & 3 & 5 \\
 5 & 76  & 92.10 & 7.89 & 3 & 4\\
6 & 68 & 89.70   & 10.29 & 8 & 1 \\
 \hline
\end{tabular}
 \caption{\label{tab:table9} Success and Failure rate of three stories in Familiar Vocabulary  }
\end{table*}

Our study showed that following a spaced repetition schedule, passphrases as natural English sentences, based on familiar vocabulary performed similarly to system-assigned passphrases based on random common words. 
However, we believe that system-assigned passphrases based on familiar vocabulary has a potential to be explored further as a new authentication method. 

Our current methodology has its limitations. The length of the passphrase in the random passphrase condition was less than the familiar vocabulary condition. To be able to construct passphrases that resembled natural language sentences from familiar vocabularies, we had to include prepositions and connecting words, resulting in passphrases of five, six and seven words. Also, passphrases less than four, from the three stories selected could not capture the style in which the story was written. We wanted to ensure that the passphrases generated from the stories not only are familiar in context of its vocabulary, but also its style. The random passphrase was only four words long as it did not contain propositions or any connection words. It is also not necessary that the chosen story was the one the participants are most familiar with and it need not be the most recent vocabulary of the participants which could have also impacted the results. For our study we are also forced to choose books whose scripts are available for free downloads. Gutenberg not only give the statistics of how many times a book was downloaded, but also give its script for free download.  Even though we may get the most popular books from Amazon, the corresponding free download may not be available of that book.  

What remains to be compared is memorability of equal length passphrases from familiar vocabulary versus random words; for example, we could remove prepositions and stop words from familiar vocabulary sentences although this comparison would fall beyond the hypothesis of this paper as the resulting passphrases will not look like natural English sentences. Another limitation to our methodology is the difficulty to compare passphrases of equal bits of security. Since the passphrase in the random vocabulary conditions is of length four and is chosen randomly from a 181-word dictionary, it had an entropy of 30 bits. But, for the three stories selected, the maximum Marginal guesswork in bits obtained for 5-grams is 14.7 bits for Alice in Wonderland, 16.8 bits for Pride and Prejudice and 16.5 bits for The Adventures of Sherlock Holmes. To even attain a 20-bit marginal guesswork, we will need to combine at least ten books  that can together generate 900,000 or more 5-grams. How to collect rich user familiar vocabularies needs further investigation. We chose three familiar stories to generate familiar vocabularies only for the sole purpose of conducting our user study as a proof-of-concept.

%%--------------------------------------------------------------------------------------------------
%%--------------------------------------------------------------------------------------------------

\bibliographystyle{IEEEtran}
\bibliography{IEEEabrv,IEEEexample}

\end{document}